\documentstyle[aps, preprint, pre, graphicx, tighten]{revtex}
\begin{document}
\title{Correlation between sequence hydrophobicity and surface-exposure
pattern of database proteins}

\author{Susanne Moelbert$^{1,2,*}$, Eldon Emberly$^{1,3}$ and Chao
Tang$^{1}$ }

\address{$^1$NEC Laboratories America, Princeton, New Jersey 08540,
USA.\\ $^2$Institut de Physique Th\'eorique, Universit\'e de Lausanne,
1015 Lausanne, Switzerland.\\ $^3$Center for Studies in Physics and
Biology Rockefeller University New York, New York 10021, USA.\\$^*$
email: susanne.moelbert@ipt.unil.ch}




\date{\today}
\maketitle

keywords: hydrophobicity, protein folding, surface exposure, secondary
structure, designability.

\newpage

\section*{Abstract}
Hydrophobicity is thought to be one of the primary forces driving the
folding of proteins. On average, hydrophobic residues occur
preferentially in the core, whereas polar residues tends to occur at
the surface of a folded protein. By analyzing the known protein
structures, we quantify the degree to which the hydrophobicity
sequence of a protein correlates with its pattern of surface exposure.
We have assessed the statistical significance of this correlation for
several hydrophobicity scales in the literature, and find that the
computed correlations are significant but far from optimal. We show
that this less than optimal correlation arises primarily from the
large degree of mutations that naturally occurring proteins can
tolerate. Lesser effects are due in part to forces other than
hydrophobicity and we quantify this by analyzing the surface exposure
distributions of all amino acids. Lastly we show that our database
findings are consistent with those found from an off-lattice
hydrophobic-polar model of protein folding.

\newpage

\section{Introduction}

One of the most persistent challenges in modern molecular biology is
to understand how proteins fold into their unique
conformation (Anfinsen 1973). The challenge lies in the fact that
there are a variety of forces which contribute to the folding process
and that these act over a range of length scales. Despite the many
interactions, it is known that a wide variety of different protein
sequences can adopt very similar folds. Analysis of the over
$20\,000$ known structures in the Protein Data Bank (PDB)
resulted in only a few hundred different folds (Murzin et al. 1995).
Although the number of determined sequences and structures increases
rapidly, the number of ``new folds'' increases only slowly, which
suggests that the total number of possible structures is extremely
small (Chothia 1992). What leads to this many to one mapping of
sequence to structure?

Of the many forces involved, it is argued that the hydrophobic
interaction plays a central role in determining the overall fold of a
protein (Kauzmann 1959, Tanford 1978). Each of the $20$ amino-acids
has a characteristic hydrophobicity -- a measure of the non-polarity
(insolubility in water) of a molecule. On average hydrophobic residues
tend to be in the core of a protein where solvent accessibility is
low, while polar residues tend to reside on the surface, where solvent
accessibility is high (Rose et al. 1985; Lesser and Rose 1990; Miller
et al. 1987; Lins et al. 2003). Many attempts based on different
approaches have been made to determine the hydrophobicity of the
amino-acids (Nozaki and Tanford 1971; Kyte and Doolittle 1982;
Engelman et al. 1986; Nautichel and Somorjai 1994;
Miyazawa and Jernigan 1996, 1999; Devido et al. 1998; Branden and
Tooze 1999).  However, the various scales in the literature sometimes
disagree as to these hydrophobicity rankings (Nautichel and Somorjai
1994) which has been attributed to the fact that hydrophobicity is a
relative quantity, which depends on the environment and reference
molecules used in the measurement (DeVido et al. 1998). Empirical
hydrophobicity measurements may not truly reflect the energetics
of solvation in protein folding (Lee 1993). Statistical scales may
reflect better the role of solvation in folding.

Although on average there is a correlation between hydrophobicity and
surface exposure (Chothia 1974; Rose et al. 1985; Miller et al. 1987),
the extent to which a fold of a protein, and hence its specific
surface-exposure pattern correlates with the hydrophobic pattern
dictated by its amino-acid sequence remains unclear.  If the average
hydrophobic behavior of amino-acids is generally true one might expect
that there should be a statistically significant correlation between
the hydrophobicity sequence and the corresponding surface-exposure
pattern. However theoretical studies of protein folding using only
hydrophobicity models (Dill 1985; Lau and Dill 1989) have shown that
there can be significant variations between hydrophobic-polar
sequences that adopt a given structure (Li et al. 1998). This
translates into the theoretical structures having a large degree of
mutational stability (Li et al. 1996). Do real proteins also display
this behavior?  Quantifying the degree of variation between sequence
and structure will be relevant to protein design based purely on
hydrophobic-polar (HP) patterning, where the hydrophobicity sequence
is assumed to dictate the final fold (Kamtekar et al. 1993).

In this article we analyze on a structure-to-structure basis the
correlation between hydrophobicity sequence and surface exposure
pattern for several commonly used hydrophobicity scales.  We find that
all the scales yield similar distributions of correlation
coefficients, and that these distributions are statistically
significant when compared to a null model in which the amino-acid
sequences are randomized. However the distributions are broad, and the
means are far from the fully correlated limit. We explore various
factors that influence this less than optimal correlation between
sequence and surface exposure pattern. This encompasses looking at how
the degree of mutational stability (i.e. sequence
entropy/designability) affects the correlation, along with other
lesser effects such as the actual surface exposure propensities of the
amino acids and secondary structural influences. We show that the less
than optimal correlation between sequence and structure for naturally
occurring proteins is a manifestation of designability, and may also be
selected for in order to 'design out' competing folds.

\section{Results}

\subsection{Testing Hydrophobicity Scales}
\label{secH-A}

In this section we compute the correlation coefficient between the
hydrophobicity sequence and surface-exposure pattern of $3242$
representative protein folds (see Methods), where the hydrophobicities
of the amino-acids are taken from several widely used hydrophobicity
scales. The scales that we have chosen to analyze are based on
different approaches: measurements of water-vapor transfer free
energies and analysis of side-chain distributions (Kyte and Doolittle
1982), semi-theoretical approaches determining transfer free energies
for $\alpha$-helical amino-acid side chains from water to a
non-aqueous environment (Engelman et al. 1986), determination of
transfer free energies by measuring solubilities in water and ethanol
relative to the reference amino-acid Glycine (Nozaki and Tanford 1971),
calculating residue-residue potentials with pairwise contact energies
(Miyazawa and Jernigan 1996), and a refined study of the latter using
the Bethe approximation for determination of relative contact energies
with respect to the native state (Miyazawa and Jernigan 1999). These
scales cover a broad range of methods used to characterize
hydrophobicity, ranging from empirical to statistical approaches.

\begin{center}
\begin{figure}[t!]
\includegraphics[bb = 0 0 792 612,width=0.95\textwidth,clip]{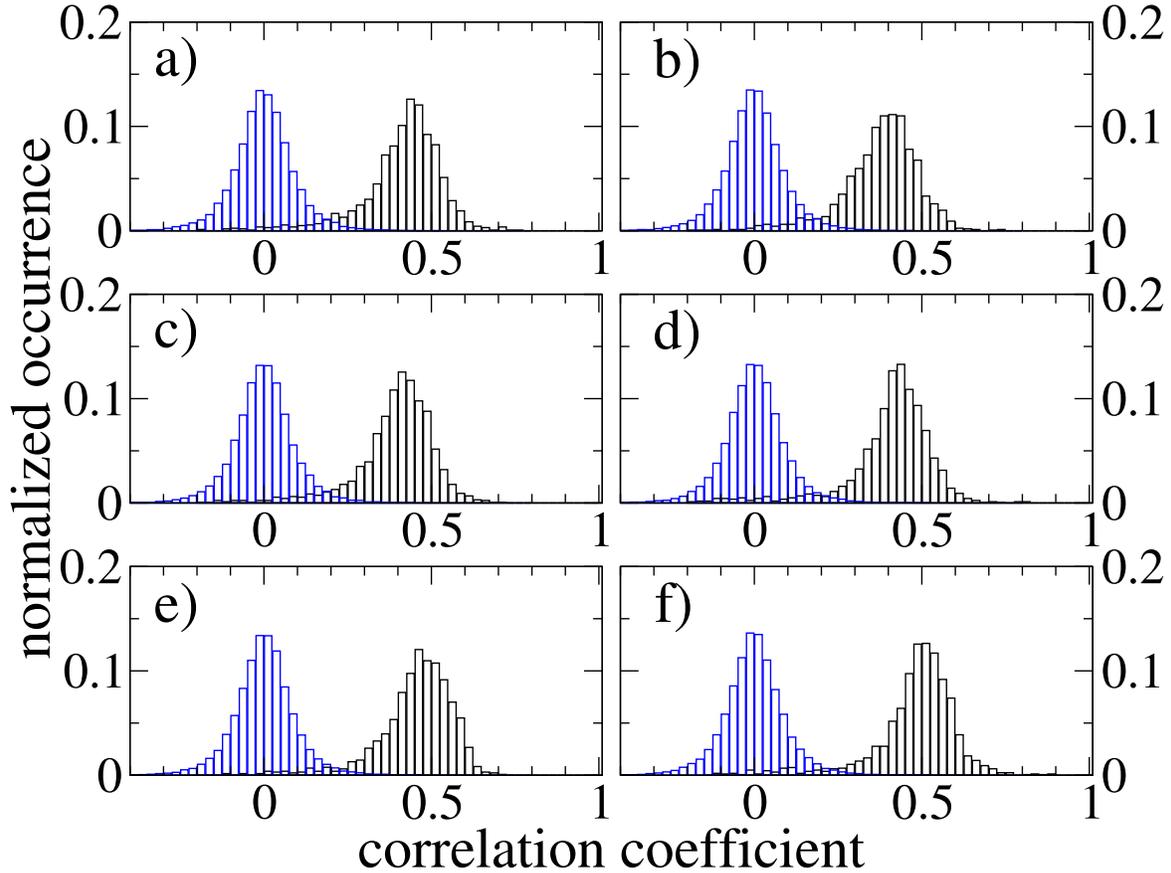}
\vspace{0.5in}
\caption{Histograms of correlation coefficients between single
surface-exposure sequences and  hydrophobicity sequences for
the $3242$ representative structures obtained using the following
hydrophobicity scales, a) Kyte and Doolittle (1982), b) Engelman et al. 
(1986), c) Nozaki and Tanford (1993), d) Miyazawa and Jernigan (1996) and e)
Miyazawa and Jernigan (1999), and f) ASA. (black bars) all amino-acids, 
(red bars) ILFVRENQ. Also shown are the histograms for the correlation 
coefficients of random amino-acid sequences (blue bars). The average 
correlation coefficients and the $Z$ scores are a) $\mu_{\rm data}=0.421$, 
$Z=2.7$, b) $\mu_{\rm data}=0.384$, $Z=2.46$, c) $\mu_{\rm data}=0.384$, 
$Z=2.46$, d) $\mu_{\rm data}=0.397$, $Z=2.55$, e) $\mu_{\rm data}=0.454$, 
$Z=2.91$, and f) $\mu_{\rm data}=0.492$, $Z=3.15$. }
\label{Fig1}
\end{figure}
\end{center}

Fig.~\ref{Fig1} shows the distributions of computed correlations
between the hydrophobicity sequences and surface-exposure patterns of
the $3242$ structures in our dataset using the above scales. The black
histograms were computed using all the amino-acids. None of the means
exceed $0.5$, with the highest being $\mu_{\rm data}=\langle
c^s\rangle_{\rm database}=0.454$ for the scale in Miyazawa and
Jernigan (1999). Nevertheless, the computed distributions are
significantly different from the null model which considers the same
set of structures but uses randomized versions of their amino-acid
sequences .  (For each representative structure we computed the
correlation coefficient between its surface-exposure pattern and $25$
random versions of its hydrophobicity sequence). The distribution of
correlation coefficients computed for the null model is shown in blue
for each scale. Despite several discrepancies in classification
between the scales, it can be seen that all yield similar
distributions of correlation coefficients and that all have similar
scores $Z = (\mu_{\rm data} - \mu_{\rm null})/\sigma_{\rm null}$ when
compared to their null models, with values between $Z=2.46$ and
$Z=2.91$ (see Table~\ref{Tab1}).

\begin{center}
\begin{figure}[t!]
\includegraphics[bb = 0 0 792 612,width=0.95\textwidth,clip]{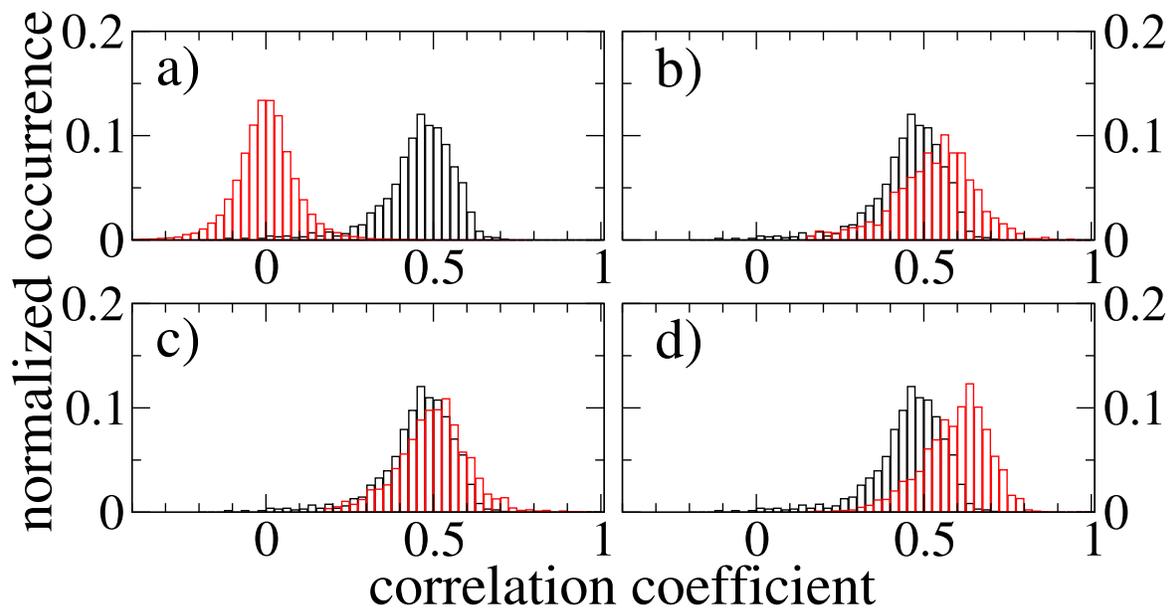}
\vspace{0.5in}
\caption{ Correlation between hydrophobicity sequence and surface
exposure for the $3242$ representative structures using the scale of
Miyazawa and Jernigan (1999) as a function of different factors. (a) No
sequence averaging (black), randomized data (red). (b) Subset of amino
acids (ILVFRENQ) (red), all amino acids (black) (c) only secondary
structure (red), whole proteins (black), (d) average over sequences
that adopt the same fold (red), no averaging (black). }
\label{Fig2}
\end{figure}
\end{center}

The above results show that a protein fold's hydrophobicity sequence
and its surface exposure pattern are far from being completely
correlated. We now explore potential reasons for this finding. In
Fig.~\ref{Fig2} we show that the correlation between hydrophobicity
sequence and surface area pattern can be improved if limits are placed
on either the sequence or the structure. For each representative
structure, we have a set of aligned structures whose sequences also
adopt the same/similar fold (see Methods). From these sequences and
structures we are able to compute an average hydrophobicity sequence
and surface exposure pattern. We find a significant improvement in the
computed correlation coefficients if the average sequences and surface
patterns are used (Fig.~\ref{Fig2}(d) and Table~\ref{Tab1}). Using
averaging over sequences to help improve structural predictions was
suggested by Finkelstein (Finkelstein 1998) and later shown
theoretically for an HP model(Cui and Wong 2000). In both those papers
it was argued that averaging was helping to reduce the noise in the
energy parameter set. With respect to sequence-structure correlation,
by averaging, one is reducing the noise contributed by sites that are
not essential to dictating the final fold. The poor correlation seen
at the single sequence level is evidence of naturally occurring
proteins having significant mutational stability or designability. We
discuss this further in the context of a model below.

A second contributing factor is that there are amino-acids for which
hydrophobicity is not the prime factor in determining exposure: as
examples, amino-acids such as glycine can appear either on the surface
or in the core, and charged amino-acids can form salt
bridges. Including such amino-acids can only lessen the correlation
between hydrophobicity and surface exposure.  We find that further
statistical significance can be achieved if only a subset of the most
hydrophobic and polar amino-acids is chosen. We have found that taking
the set of amino-acids [ILFVRENQ] results in an appreciable
improvement in the $Z$ score (Fig.~\ref{Fig2}(b) and
Table~\ref{Tab1}). The four hydrophobic residues were chosen because
they are the largest, and adding others reduced $Z$.  The four polar
residues were selected because they have the largest ratio of polar
surface area to hydrophobic surface area. Hence including those
amino-acids for which hydrophobicity is most likely to be the dominant
force in determining their surface exposure within a protein fold
indeed improves the correlation. In the next section we explore in
much more detail the propensities for surface exposure of each of the
amino acids.

Lastly, we consider the improvements to the correlation between
sequence and structure if only residues that form secondary structural
elements are used. Many helices and strands have one side that is
hydrophobic and hence tends to be in the core while the other side is
polar and tends to be exposed on the surface. Turns tend to be
flexible and irregular. Including turns may increase the noise in the
data. Fig.~\ref{Fig2}(c) shows that a slight improvement is gained by
only considering helices and strands.  We will further break down the
connection to secondary structural elements and surface exposure for
the various amino acids below.

\subsection{Surface-Exposure Distributions of the Amino-Acids}
\label{secASA}

As shown in Sec.~\ref{secH-A}, the known hydrophobicity scales yield
statistically significant correlations between a protein's pattern of
surface exposure and the hydrophobicities of its amino-acid
sequence. However, despite this statistical significance, the
correlations are far from the case where hydrophobicity and exposure
patterns are completely correlated. In this section, we show that this
departure from optimal correlation can be partly attributed to the
broad distribution of surface exposures which some amino-acids
tolerate. In the spirit of the work by Rose et al. (Rose et. al 1985),
for each amino-acid we have computed its surface exposure distribution
within the representative set of structures. From the distributions we
derive a surface-exposure propensity that reflects the tendency of
each amino-acid to be either exposed or buried in the core, and show
that this scale leads to a better correlation between sequence and
surface pattern.

\begin{center}
\begin{figure}[t!]
\includegraphics[bb = 0 0 792 612,width=0.95\textwidth,clip]{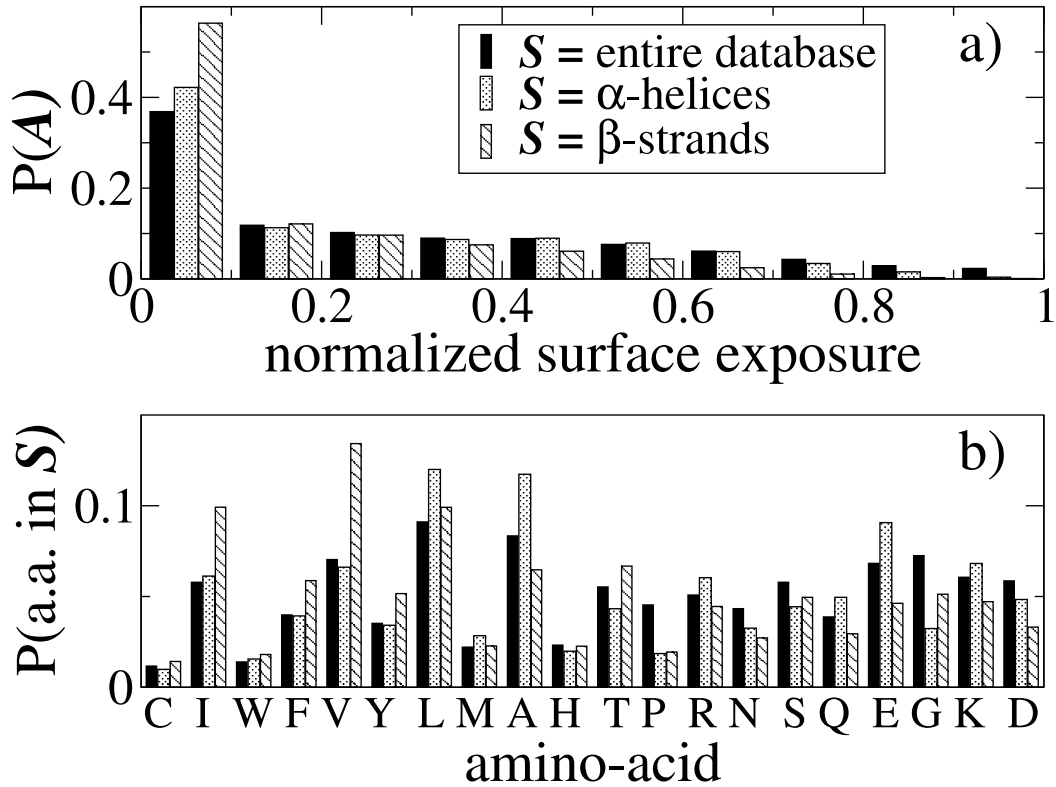}
\vspace{0.5in}
\caption{ (a) Probability of finding a residue at a given degree of
surface exposure ${\cal A}$ (${\cal A}=0$: core, ${\cal A}=1$:
surface) compared to the probability of finding an $\alpha$-helix
residue and a $\beta$-strand residue at a given degree of surface 
exposure ${\cal A}$.  (b) Probability of finding a residue in an
$\alpha$-helix and in a $\beta$-strand compared to the probability of
finding it at any position in a protein. The total number of residues
in proteins is $352\,707$, in $\alpha$-helices $129\,643$, and in
$\beta$-strands $74\,543$. }
\label{Fig3}
\end{figure}
\end{center}

Before considering the surface-exposure distributions of each amino-acid, 
we examine the probability distributions for surface exposure
and amino-acid occurrence within the database of structures. Folded
proteins are dense, three-dimensional clusters of amino-acids. The
core thus represents a considerable portion of the whole protein,
whereas only a relatively small number of amino-acids are to some
extent exposed to the aqueous solvent. In Fig.~\ref{Fig3}(a) we 
show the probability $p(\mathcal{A})$ for a given surface exposure
$\mathcal{A}$ using all of the side-chain exposures from the $3242$
representative structures. It is clear that a large fraction of
residues reside in the core, where surface exposure is low. The
probability of occurrence for the individual amino-acids, $p(a.a.)$, is
also non-uniform. Fig.~\ref{Fig3}(b) shows the occurrence 
frequencies of the amino-acids within the sequences used in the dataset.
These distributions will be used to examine whether the occurrence of
an amino acid with a given surface exposure is correlated or independent.

For each amino-acid we compute the joint probability of observing a
given surface exposure, $p(a.a. \& \mathcal{A})$. In order to extend
the analysis of Rose et al. and to better characterize the propensity
of a given amino-acid to appear with a given surface exposure, we
compare the joint probability with the null model where the occurrence
of an amino-acid and the surface exposure are independent. This is
expressed by the ratio,
\begin{equation}
{\mathcal{P}} = \frac{p(a.a. \& {\mathcal{A}})}{p(a.a.)
p({\mathcal{A}})}, \label{eqPdistrib}
\end{equation}
where values greater than one indicate favored for the given surface exposure,
while those less than one are less favored.

\begin{center}
\begin{figure}[t!]
\includegraphics[bb = 0 0 792 612,width=0.95\textwidth,clip]{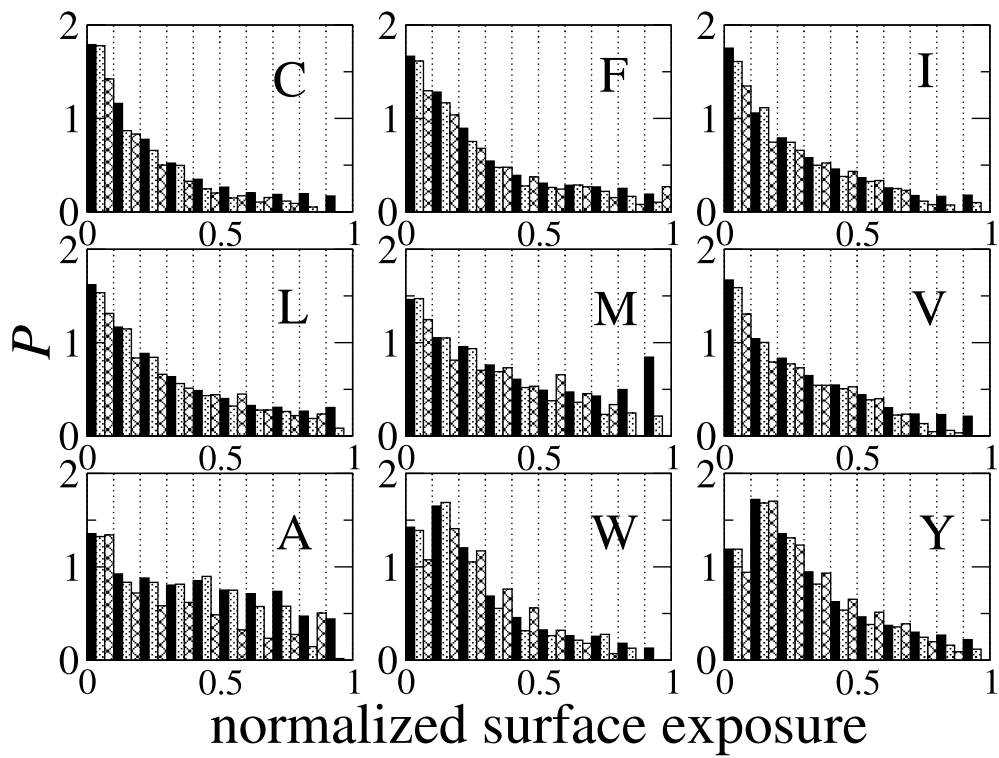}
\vspace{0.5in}
\caption{Histograms of degree of surface exposure of the core
amino-acids ($\mathcal{C}$) in the complete database, only in
$\alpha$-helices, and only in $\beta$-strands. Legend as in
Fig.~\ref{Fig4}.}
\label{Fig4}
\end{figure}
\end{center}

\begin{center}
\begin{figure}[t!]
\includegraphics[bb = 0 0 792 612,width=0.95\textwidth,clip]{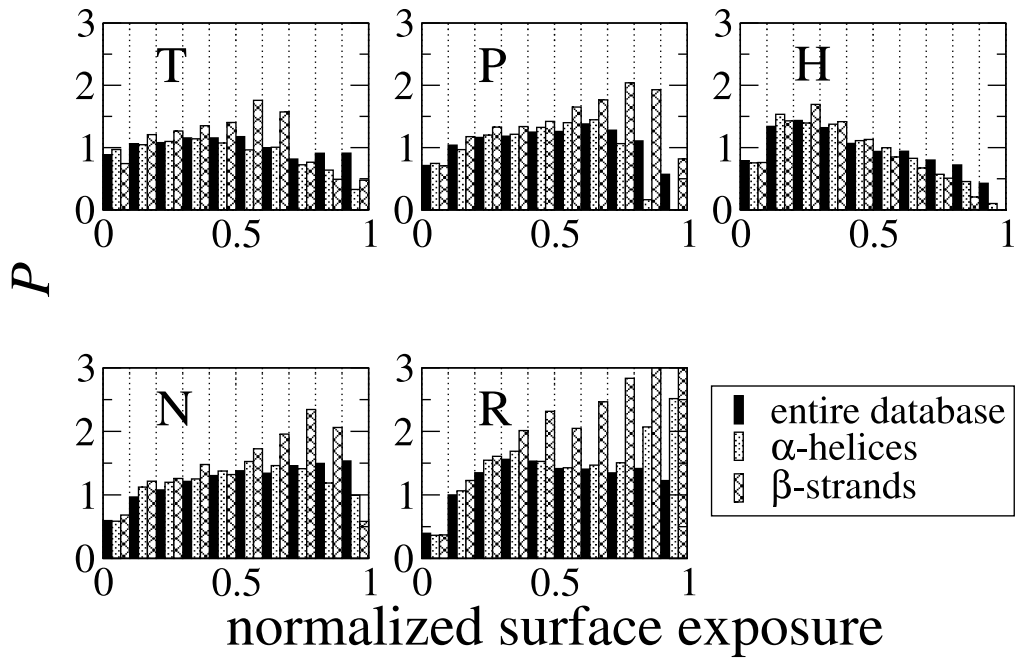}
\vspace{0.5in}
\caption{Histograms of degree of surface exposure of the intermediate
amino-acids ($\mathcal{M}$) in the entire database, only in
$\alpha$-helices, and only in $\beta$-strands.}
\label{Fig5}
\end{figure}
\end{center}

\begin{center}
\begin{figure}[t!]
\includegraphics[bb = 0 0 792 612,width=0.95\textwidth,clip]{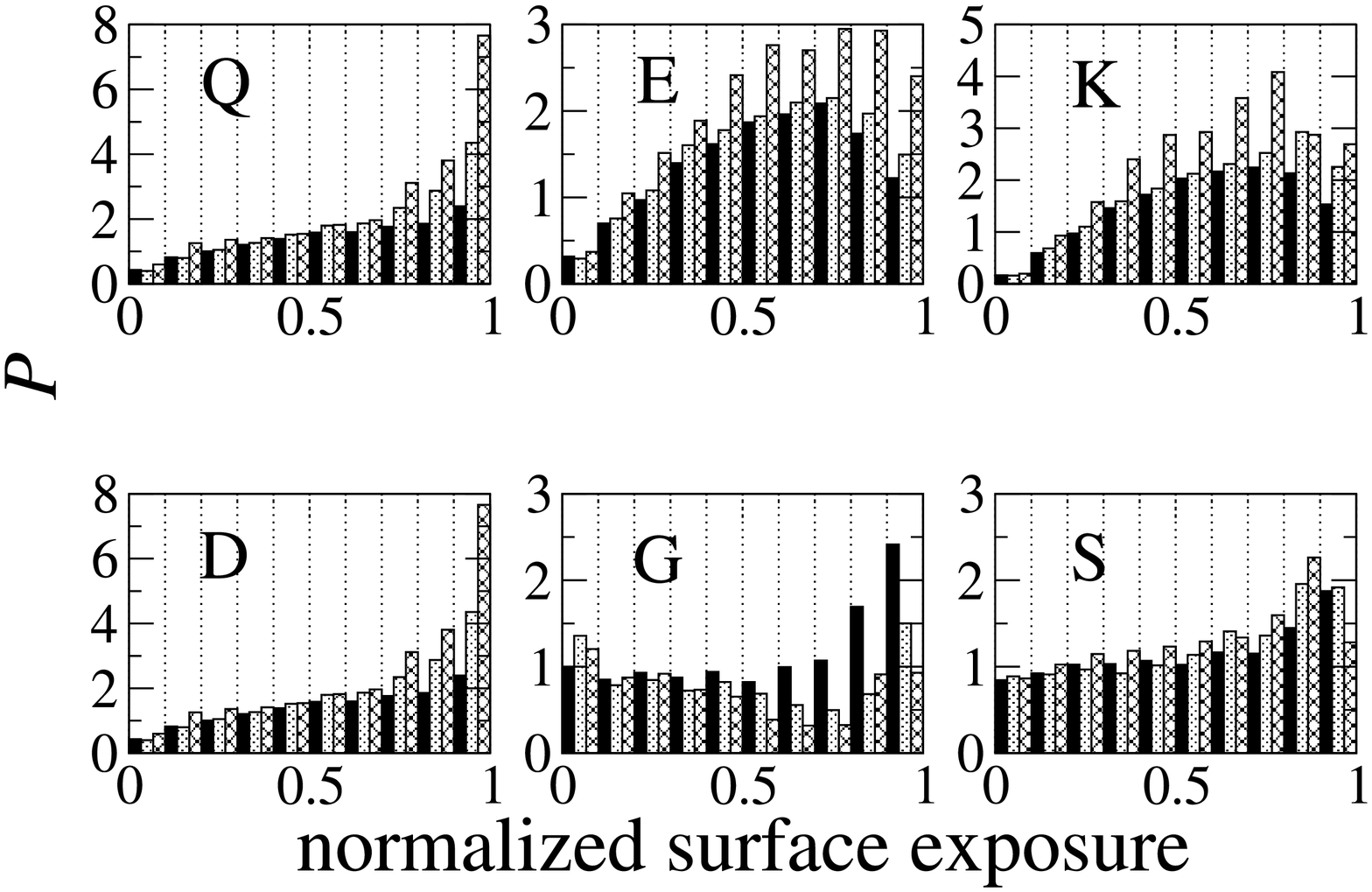}
\vspace{0.5in}
\caption{Histograms of degree of surface exposure of the surface
amino-acids ($\mathcal{S}$) in the complete database, only in
$\alpha$-helices, and only in $\beta$-strands. Legend as in
Fig.~\ref{Fig4}.}
\label{Fig6}
\end{figure}
\end{center}

Figs.~\ref{Fig4}-\ref{Fig6} show the distributions of $\mathcal{P}$
for the $20$ amino-acids. The distributions are rather broad. Tests
using only a half of the database, and others using only a half of the
length of the sequences, led to very similar results. As was found by
Rose et al., our distributions are also suggestive of three classes of
amino-acids: core amino-acids ($\mathcal{C}$) with a peak at low
surface exposure, surface amino-acids ($\mathcal{S}$) with a peak at
high surface exposure, and intermediate amino-acids ($\mathcal{M}$)
with relatively flat distributions. We are in agreement with Rose
et. al regarding core amino-acids, however there are discrepancies
between our classification of intermediate and surface amino
acids. Nominally some of our intermediate amino acids show preferences
for being on the surface when only secondary structure is considered
-- this is discussed below. 

For each amino-acid, the mean of the $\mathcal{P}$ distribution gives
a weighted average surface accessibility (ASA) for each
amino-acid. Tab.~\ref{Tab2} shows the computed ASAs of the $20$
amino-acids. Although the surface-exposure scale ranges from $0$
(completely hidden in the core) to $1$ ($100\%$ exposed to water), the
averages do not take extreme values. $11$ amino-acids have rather
moderate tendencies to prefer the core of proteins, while $9$ are more
polar. Tyrosine occurs mostly in the core, and thus shows quite
hydrophobic properties in a protein environment. Charged amino-acids
including Aspartic acid, Glutamic acid, Lysine, and Arginine, not
surprisingly, tend to occur on the surface. Cysteine is the monomer
most frequently found in the core, and thus represents the most
markedly hydrophobic amino-acid. Thus despite Cysteine having a polar
group, it has a strong tendency to be buried in the core, which can be
attributed to its ability to form disulfide bonds within the cores of
protein structures. 

\begin{center}
\begin{figure}[t!]
\includegraphics[bb = 0 0 792 612,width=0.95\textwidth,clip]{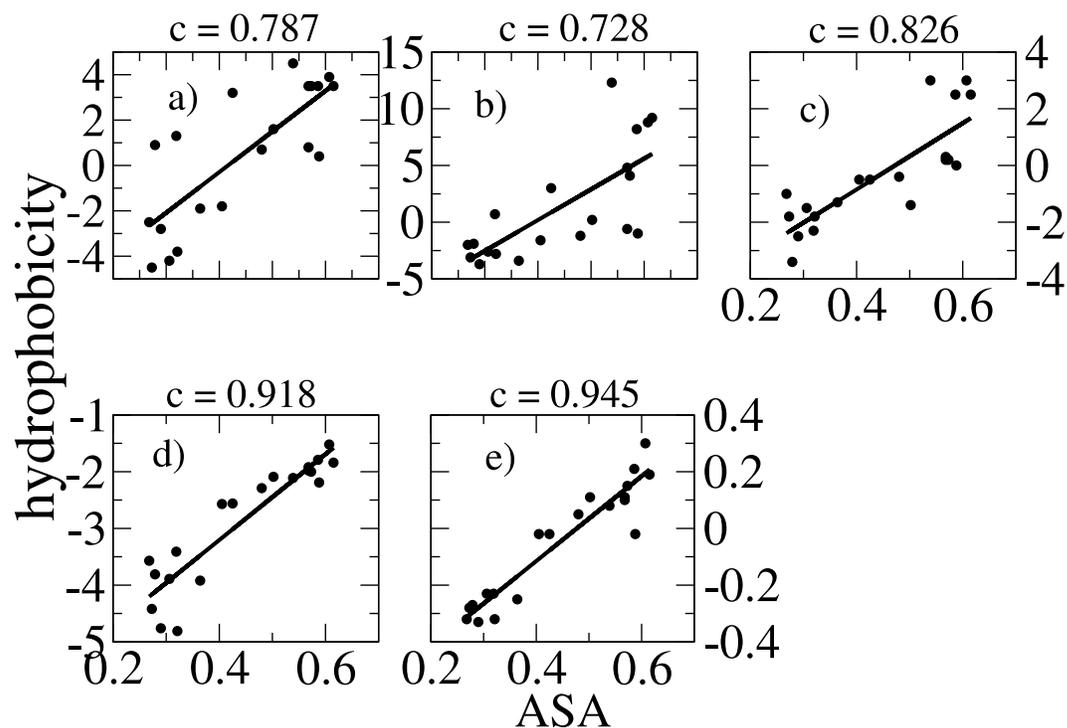}
\vspace{0.5in}
\caption{Correlation between ASA values of the $20$ amino-acids
(Tab.~\ref{Tab2}) and their hydrophobicity values deduced from the
scales of a) Kyte and Doolittle (1982), b) Engelman et al. (1986), c)
Nozaki and Tanford (1993), d) Miyazawa and Jernigan (1996), and e)
Miyazawa and Jernigan (1999).  }
\label{Fig7}
\end{figure}
\end{center}

Comparison to the hydrophobicity scales shows that the ASA scale
agrees in large part with the method of Miyazawa and Jernigan (1999)
as regards the broad distinction between hydrophobic and polar
amino-acids. However, the specific rankings are rather different.
Correlations between the ASA values for the $20$ amino-acids and their
hydrophobicity values determined using the scales under consideration
are shown in Fig.~\ref{Fig7}. The three scales based on the transfer
free energies of amino-acid side chains from water into either vapor
or non-aqueous solvents have the lowest correlation with the ASA
scale. An improvement is observed for the scales obtained by
determination of the pairwise interaction between amino-acids. Thus
the database derived hydrophobicity scales correlate best with our
statistically derived surface exposure propensities. The lesser
correlation to empirical scales highlights the context dependence of
hydrophobicity and that there are departures between how an amino acid
behaves in liquid solution versus the environment of densely packed
protein. This highlights how energetics depends on the reference state
whose affects on the correlation between a similar set of parameter
sets was discussed by Godzik et al. (1995).

We conclude this subsection by re-examining the correlation between the
amino-acid sequence and surface-exposure pattern of a protein. Using
the ASAs in Tab.~\ref{Tab2} we assign to each amino-acid sequence a
most probable surface-exposure pattern. Table~\ref{Tab1} shows the
results of the correlation analysis using this scale. These database
derived mean surface exposures for each amino acid consistently yield
 better correlation coefficients than the hydrophobicity scales. Thus
using the above surface exposure distributions to derive statistical
surface propensities may offer a better alternative to the
hydrophobicity scales that we have examined.

\subsection{Secondary-Structure Analysis}

The native configuration of a folded protein is characterized by
secondary-structure elements, $\alpha$-helices and $\beta$-strands,
which are connected by turns (Levitt and Chothia 1976). It was shown
above that considering only the sequence and surface patterns of
secondary structural elements led to a slight improvement in the
correlation between hydrophobicity and exposure. In this section we
break down the occurrence of the $20$ amino-acids in these structural
elements and their corresponding surface-exposure patterns. We first
consider the distribution of surface exposures within secondary
elements irrespective of amino-acid: Fig.~\ref{Fig3}(a) shows that
most of the residues in

$\alpha$-helices and $\beta$-strands occur in the interior of native
protein configurations. However, this effect is much stronger for
$\beta$-strands indicating that residues making up $\beta$-strands have
a higher tendency to be in the core than those making up helices.

It is well known that the various amino-acids have different
propensities to form either $\alpha$-helices or $\beta$-strands
(Munoz and Serrano 1994). Fig.~\ref{Fig3}(b) shows the
frequency of occurrence of each amino-acid in $\alpha$-helices and
$\beta$-strands compared to the frequency of occurrence over the whole
database.  The amino-acids are arranged according to their ASA values
in increasing order.  Compared to the total database, $\beta$-strands
tend to be composed of a high portion of amino-acids with low ASA and
rather large side chains, such as V, I, and T, or with an aromatic ring 
as in F, Y, and W, while charged amino-acids occur less frequently than
expected. For $\alpha$-helices, strong helix-formers such as Alanine are 
particularly prominent, and the residues which are found more
frequently in other parts of the proteins are
divided into comparable numbers of amino-acids with low and high
ASA.

Figs.~\ref{Fig4}-\ref{Fig6} show the surface-exposure 
distributions $\mathcal{P}$ of the $20$ amino-acids in $\alpha$-helices 
and in $\beta$-strands, juxtaposed with the distributions for the entire
database. For the core ($\mathcal{C}$) amino-acids, the differences are 
rather small. However, for the intermediate ($\mathcal{M}$) amino-acids, both
Arginine and Asparagine (which are nominally polar) appear prominently
as being exposed in $\beta$-strands. Arginine is also seen to have a
tendency to appear on the exposed surfaces of helices. For those nominally 
polar amino-acids ($\mathcal{S}$) classified as residing on the surface, 
the propensity to be exposed is further increased within secondary
structures when compared to the results obtained from the whole
database. These slight enhancements in surface exposure propensity for
certain amino acids while in secondary structural elements leads to
the marginal improvement in correlation between sequence and surface
exposure seen above when only secondary elements were included.

\subsection{Model}

Theoretically, hydrophobic-polar (HP) models have been studied for
some time to help clarify the nature of the hydrophobic force in the
folding process. Correlations have been studied in the context of
sequence (White \& Jacobs 1990), and non-randomness has been detected
both in real protein sequences and theoretical models (Irb{\" a}ck et
al. 1996; Irb{\"a}ck and Sandelin 2000).  Here, we consider the
correlations between hydrophobicity sequences and surface-exposure
patterns which emerge in a protein-folding model based solely on
hydrophobicity. Does the less than perfect correlation between
hydrophobicity sequence and surface pattern still remain when only
solvation energy is considered? If so, is it due to the large
variation of sequences that can be tolerated by highly designable
structures (Li et al. 1996)? How does averaging improve the
correlation in the model results?

We study the folding of random amino-acid sequences using an HP model
(see Methods), where the single energy entering the analysis is a
solvation energy dependent only on the hydrophobicities of the side
chains and their corresponding surface exposures in a fold. It is not
computationally feasible to consider the continuum of possible
structures which a large set of random sequence could adopt, so we
choose to use only a finite number of compact
representative folds, formed in this case by a statistically complete
set of four-helix bundles. The designability of this set of structures
has been studied previously, and many of the top designable helix
structures in this set correspond to naturally occurring four-helix
bundles (Emberly et al. 2002). The set has the following advantages:
(a) the folds are 60mers and hence are much longer than structures
generated by enumerating all possible structures using a finite set of
dihedral angles (Miller et al. 2002); (b) it is more diverse than
decoy sets generated from a specific native fold. A set of random
amino-acid sequences was folded onto the above set of structures using
the HP model (see Methods). We chose the top 250 designable structures
and their corresponding sequences to form the database on which to
perform the correlation analysis. These structures represent plausibly
thermodynamically stable folds and their corresponding sequences,
although just a mere sample of the sequences that actually fold into
these structures, are assumed to be good folders. Lattice studies have
shown that removing the compactness constraint can lead to a different
set of designable structures (Chan and Bornberg-Bauer 2002), but the
correlation findings below undoubtedly would not change.

\begin{center}
\begin{figure}[t!]
\includegraphics[bb = 30 30 575 700,width=0.75\textwidth,clip,angle=270]{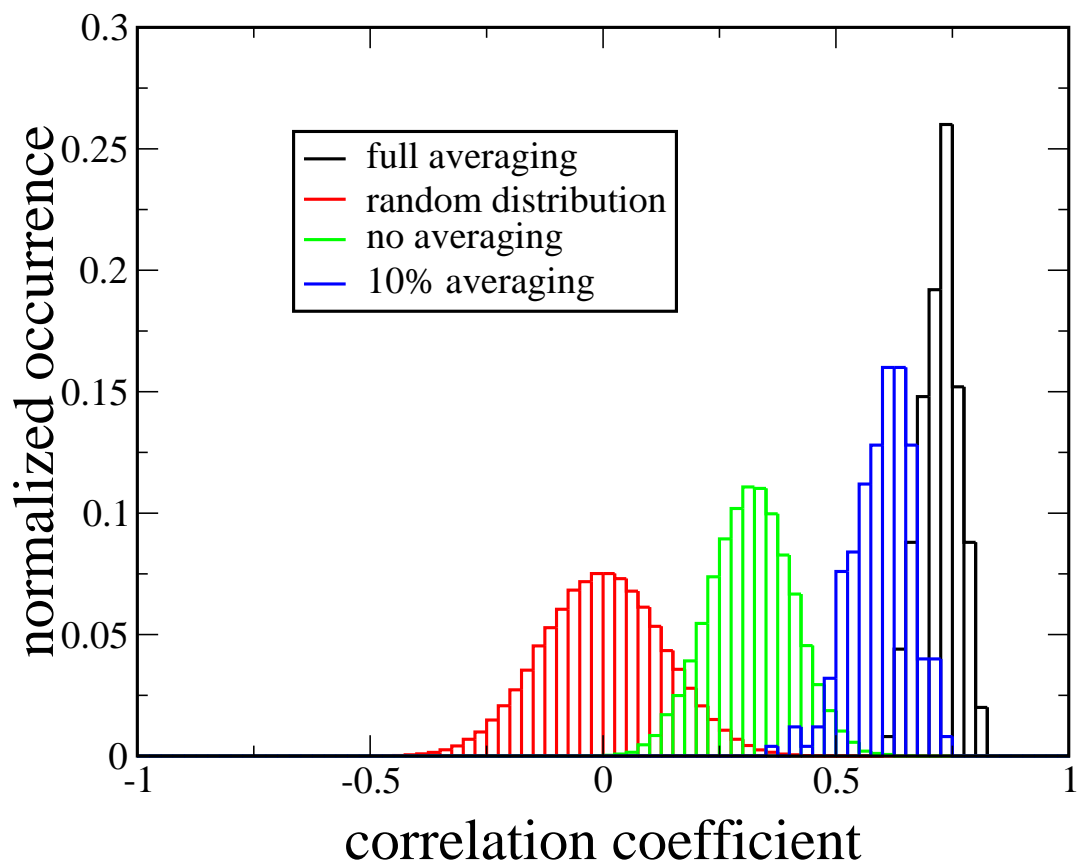}
\vspace{0.5in}
\caption{Histograms of correlation coefficient computed for the
average hydrophobicity sequences and surface-exposure patterns of the
top 250 designable model four helix bundles (black). The distribution
of correlation coefficients for the null model where the sequences
were randomized is shown in red. The scale of Nozaki and Tanford
(1971) was used. }
\label{Fig8}
\end{figure}
\end{center}

Fig.~\ref{Fig8} shows the distribution of correlation coefficients
between the hydrophobicity sequences and surface-exposure patterns of
the model. The green histogram was computed using only a single
sequence, randomly selected from the pool that fold to the corresponding
structure, for each structure. This is nearly identical to what was found
from the database, namely that the correlation between a
hydrophobicity sequence and its structure is less than optimal. The
red histogram is for a randomized version of the data. Thus, as before
the correlation between sequence and structure is not random and has
some statistical significance. Because for each of the 250 designable
structures we have several hundred sequences which fold into them, we
can assess the effects of sampling. As in the analysis for the real
protein structures, the mean hydrophobicity sequence was computed for
each set of sequences that adopt the same fold. Although the mean is
somewhat greater than those of the database distributions, the model
distribution remains similar to the results computed from the database
structures and sequences. Reducing the number of sequences used to
compute the average (10\%) still leads to an improvement in the
correlation and is more in line with the improvement seen in the
database analysis. We discuss the implications of the theoretical
findings in light of the database results below.

\section{Discussion}

Hydrophobicity has long been considered as one of the primary driving
forces in the folding of proteins. It has been shown, and reconfirmed
by our results that the hydrophobicity of an amino-acid is indeed
correlated with its average surface exposure. However, the degree to
which this correlation extends to the relationship between specific
amino-acid sequences and surface patterns has received little
investigation. We have now quantified this correlation for several
widely used hydrophobicity scales, and have shown that amino-acid
hydrophobicity does play a statistically significant role in shaping
the surface-exposure pattern of a structure. However the distributions
of correlation coefficients are broad, and remain far from the optimal
case in which the surface-exposure pattern would show a perfect
correlation with the hydrophobicity pattern.

The origin of this suboptimal correlation may lie in the fact that
there are factors other than hydrophobicity which contribute to the
determination of a protein's final fold. There are clearly other
forces at work in determining a protein's ultimate fold -- e.g. a recent
study suggested that hydrophobicity alone can not account for the observed
thermodynamics of protein folding (Chan 2000). Thus some residues
behavior may not be solely dictated by hydrophobicity. Using updated
data, we carried out an analysis similar to Rose et al. (1985) to
determine the surface-exposure distributions of each of the
amino-acids, and found that many were rather broad. Indeed, several
amino-acids have essentially flat distributions, and hence their
exposure seems to be uncorrelated with their hydrophobicity. Such
broad distributions are in part responsible for the less than optimal
correlation, and we showed that using only a subset of amino acids
which have more peaked distributions led to improved correlations.
The exposure distributions reflect all of the forces that are involved
in the folding process, and we have found several discrepancies
between the most likely exposure of an individual amino-acid and its
hydrophobicity. An example is provided by Cysteine, for which the
ability to form disulfide bonds with other Cysteine residues
constitutes a factor independent of hydrophobicity which influences
surface exposure. From the distributions we computed a scale which
reflects the surface exposure propensities of the amino acids. This
goes beyond just hydrophobicity and leads to an improvement in the
correlation between sequence and the surface exposure pattern of a
fold. Hence for folding studies that use energy models that are based
solely upon side chain solvation, using these database derived
distributions (or the ASA's computed from them) over the empirical
hydrophobicity scales should lead to a better performance.

By far the greatest improvement was achieved when we computed the
correlation coefficients between average hydrophobicity sequences and
structures. The average hydrophobicity sequence gives a better measure
of the sequence which best matches the structure (Finkelstein 1998). 
The low correlation observed at the single
sequence level shows that there can be a broad variation from that of
the ``best match'' sequence. From theoretical models it is predicted
that thermodynamically stable folds are those which are also highly
designable, that is, they have a large number of sequences that fold
into them (Li et al. 1996; Miller et al. 2002; Emberly et al. 2002).
This large degree of mutational stability for designable folds means
that there can be significant departures from the lowest energy
sequence. In fact if sequences were selected at random from a large
pool of sequences that fold into a designable structure, it would be
more likely to select a sequence far from the central ``best match''
sequence than not. Even if a sequence started near the ``center''
(best match sequence), its ``neutral'' evolution would lead it to
somewhere farther away from the center in the sequence space due to
the sequence entropy (Li et al. 1998; Taverna and Goldstein 2002a).
Hence the lack of strong correlation between sequence and structure
found in the database could be a signature of designability in
nature. It has also been postulated that it may even be advantageous
for sequences to select against being near the ``best match'' as such
selection helps to improve plasticity in sequence space (Taverna and
Goldstein 2002b).

We have shown that the correlation improves when one uses the average
hydrophobicity sequence, however we have also found that even the
average sequence is not perfectly correlated with the surface exposure
pattern. This could simply be due to insufficient sampling of sequence
space or could be evidence of something more fundamental. It has been
argued that having a suboptimal correlation between a protein's
amino-acid sequence and surface exposure pattern may help to improve
the thermodynamic stability of the fold and ``design out'' competing
folds (DeGrado 1997). All of the average hydrophobicity patterns for
the most designable model helix structures have ``misspellings'' at
various locations, where a misspelling involves the placement of a
hydrophobic residue on an exposed site or a polar residue in the
core. These departures from the optimal pairing of hydrophobicity with
exposure have been shown in other theoretical studies (Emberly et
al. 2002) to help increase the energy gap between the ground state and
competing structures. If the hypothesis of designing out competing
structures through suboptimal correlation is valid, this has important
consequences for structural design based on binary patterning
(Kamtekar et al. 1993). The surface pattern of the structure may act
as a starting point for the selection of an amino-acid sequence, but
it may then prove advantageous to depart from this blueprint in order
to improve thermodynamic stability. Database analysis of the type performed here may
form the basis for advanced techniques to detect further correlations
between sequence and structure which would help to better design sequences
in protein design.

\acknowledgments
We wish to thank Jonathan Miller and Bruce Normand for many helpful
comments and discussions. This research was partially supported by The
Swiss Study Foundation and by the Swiss National Science Foundation
through grants FNRS 21-61397.00 and 2000-67886.02.

\section{Methods and Model}
\label{SecMethModel}
\subsection*{Representative set of database structures}
In order to have a non-redundant set of protein structures for
analysis, we have chosen to use the $3242$ representative structures
from the FSSP database (Holm and Sander 1996). The FSSP database is
the result of an all-against-all structure analysis which groups
protein structures into a hierarchical tree based on their level of
structural similarity. All residues of the known protein structures
are compared in three dimensions, and the results are reported in the
form of alignments of equivalent residues. Redundancy is eliminated by
removing proteins with mutual sequence identity larger than $25\%$,
because they result in almost complete structural overlap. There are
$30624$ known protein chains grouped to one of the representative
structures in the FSSP.  Each representative structure has a set of
aligned structures. Each structure in turn has a corresponding
amino-acid sequence.  Thus for each representative structure in the
FSSP we have a list of aligned structures along with a corresponding
set of amino acids sequences, all of which are assumed to fold into a
similar fold as the representative structure in the aligned regions.

\subsection*{Correlation analysis}

A hydrophobicity scale $s$ assigns a hydrophobicity value $h^s_{{\sl
a.a.}}$ to each amino-acid ({\sl a.a.}). $h^s_{i,j}$ is the
hydrophobicity of the $i{th}$ aligned residue of sequence $j$ which is
aligned with a representative structure, based on the hydrophobicity
scale $s$. For the set of amino-acid sequences which fold into a given
structure we wish to consider what the average hydrophobicity sequence
for the set is. We consider the average sequence because it gives a
good characterization of the hydrophobicity sequence which adopts the
given representative structure (Finkelstein 1998, Cui 2000). The
average hydrophobicity value $\overline{h^s_i}$ at position $i$ within
this representative structure using scale $s$ is
\begin{equation}
\overline{h^s_i} =\frac{1}{M} \sum_{j=1}^{M} h^s_{i,j},
\end{equation}
where $M$ is the number of sequences in the alignment at residue
$i$. Calculating this average for all residues of the representative
structure with length $N$ gives the average hydrophobicity sequence of
this structure, $(\overline{h^s_i})_{i=1..N}=(\overline{
h^s_{1}},\overline{ h^s_{2}},\cdots ,\overline{ h^s_{N}})$.

The surface exposure $a_i$ of residue $i$ in a structure is quantified
as the amount of surface area of the side chain atoms (represented as
spheres) that is accessible to water (represented by a sphere of
radius 1.4 \AA). For each structure we obtain the surface exposures of
each of its residues from the FSSP file. We normalize each surface
exposure by the total surface area of the side chain atoms making up
the given residue (Creighton 1993). This yields a fractional exposure
for each residue in a structure.  We compute an average surface
exposure pattern for a structure using its FSSP alignment,
\begin{equation}
\overline{a_i} = \frac{1}{L}\sum_{j=1}^{L} a_{i,j}^{\gamma},
\label{eq:aMoy}
\end{equation}
where $L$ is the number of known structures which have a residue
aligned with residue $i$ and $a_{i,j}$ denotes the surface-accessible
area of residue $i$ in structure $j$ of the alignment. Performing this
procedure for each residue $i$ of the representative structure leads
to a sequence of surface accessibilities $(\overline{a_i})_{i=1..N} =
(\overline{a_1}, \overline{a_2}, \cdots \overline{a_N})$.

The correlation coefficient $c^s$ between the hydrophobicity
sequence $(\overline{h^s_i})_{i=1..N}$ and the accessible-surface area
sequence $(\overline{a_i})_{i=1..N}$ of a structure is given by
\begin{equation}
c^s = \frac{\sum_{i}^{N}
(\overline{a_i}-\overline{a})(\overline{h^s_i}-\overline{h^s})
}{\sqrt{\sum_{i}^{N} (\overline{a_i}-\overline{a})^2 \sum_{i}^{N}
(\overline{h^s_i}-\overline{h^s})^2 } }.
\label{eqCorrCoeff}
\end{equation}

\subsection*{Hydrophobic-Polar Model}

In hydrophobic-polar (HP) models, hydrophobicity is the sole force
driving the folding process (Dill 1985; Lau and Dill 1989). For an 
amino-acid sequence which corresponds to a sequence of hydrophobicities 
$\{h_i\}$, the solvation energy of the sequence on a given structure 
$\gamma$ is,
\begin{equation}
E^{\gamma} =  \sum_{i=1}^N h_{i} (1 - a_{i}^{\gamma})
\end{equation}
where $a_i^\gamma$ is the surface exposure of residue $i$ in structure
$\gamma$. The native fold of a sequence is the one which minimizes this
energy.

We use a representative set of structures to act as the space of
potential folds. For a given amino-acid sequence we then use the above
energy equation to determine the structure which has the lowest
energy within the set of competing structures. We deem this to be the
native fold of the sequence. Studies have shown that folding numerous
random amino-acid sequences in this way results in a non-uniform
mapping of sequences to structures: some structures turn out to be native 
folds far more often than others, and have been designated ``designable''
structures (Li et al. 1996). 

Here we consider a representative set of 203282 four-helix bundles for
the competing set of structures (Emberly et al. 2002). This set was
shown to cover the space of all possible four-helix folds at the 95\%
confidence level, and hence represents a relatively complete set of
compact folds on which a HP sequence can compete. 10$^6$ random
amino-acid sequences (the hydrophobicity scale based on transfer free
energy between water and ethanol was used (Nozaki and Tanford 1993)), were
folded by selecting the ground-state structure for each sequence. The
top 250 designable structures (each with several hundred sequences
which fold into it) and their corresponding hydrophobicity sequences
formed the model database on which the correlation analysis was
performed.

\section*{References}

\noindent Anfinsen, C.B. 1973. Principles that govern the folding of
protein chains. {\sl Science} {\bf 181:} 223-230.\\

\noindent Branden, C. and Tooze, J. 1999. {\sl Introduction to Protein
Structure,} pp. 6-7. Garland Publishing, New York. \\

\noindent Chan, H.S. 2000. Modeling Protein Density of States: Additive
Hydrophobic Effects are Insufficient for Calorimetric Two-State
Cooperativity. {\sl Proteins} {\bf 40:} 543-571. \\

\noindent Chan, H.S. and Bornberg-Bauer, E. 2002.  Perspectives on
protein evolution from simple exact models. {\sl App. Bioinformatics}
{\bf 1:} 121-144. \\

\noindent Chothia, C. 1974. Hydrophobic bonding and accessible surface
area in proteins. {\sl Nature} {\bf 248:} 338-339.\\

\noindent Chothia, C. 1992. One thousand families for the molecular
biologist. {\sl Nature} {\bf 357:} 543-544.\\

\noindent Creighton, T.E. 1993. {\sl Proteins: Structures and
Molecular Principles,} 2nd ed., W.H. Freeman, New York. Hydrophobicity
scale, p. 154. Surface accessibilities of amino acids, p.142. \\

\noindent Cui, Y., and Wong., W.H. 2000. Multiple-Sequence Information
Provides Protection against Mis-Specified Potential Energy Functions
in the Lattice Model of Proteins. {\sl Phys. Rev. Lett} {\bf 85:} 5242-5245.\\

\noindent Dill, K.A. 1985. Theory for the folding and stability of
globular proteins. {\sl Biochemistry} {\bf 24:} 1501-1509; \\

\noindent DeGrado, W. F. 1997. Proteins from scratch. {\sl Science}
{\bf 278:} 80-81.  \\

\noindent DeVido, D.R., Dorsey, J.D., Chan, H.S., and Dill,
K.A. 1998. Oil/Water Portioning Has a Different Thermodynamic
Signature When the Oil Solvent Chains Are Aligned Than When They Are
Amorphous. {\sl J. Phys. Chem.} {\bf 102:} 7272-7279. \\

\noindent Emberly, E.G., Wingreen, N.S., and Tang,
C. 2002. Designability of $\alpha$-helical proteins. {\sl PNAS} {\bf
99:} 11163-11168.\\

\noindent Engelman, D.M., Steitz, T.A., and Goldman
A. 1986. Identifying nonpolar transbilayer helices in amino-acid
sequences of membrane proteins. {\sl
Annu. Rev. Biophys. Biomol. Struct.} {\bf 15:} 321-353. \\

\noindent Finkelstein, A.V. 1998. 3D Protein Folds: Homologs Against
Errors--a Simple Estimate Based on the Random Energy Model. {\sl
  Phys. Rev. Lett.} {\bf 80:} 4823-4825. \\

\noindent Godzik, A., Kolinski, A. and Skolnick J. (1995) Are Proteins
Ideal Mixtures of Amino Acids? Analysis of Energy Parameter
Sets. {\sl Prot. Sci.} {\bf 4:} 2107-2117. \\

\noindent Holm, L. and Sander, C. 1996. Mapping the protein
universe. {\sl Science} {\bf 273}:595-602.  \\

\noindent Irb{\" a}ck, A., Peterson, C. and Potthast,
F. 1996. Evidence for nonrandom hydrophobicity structures in protein
chains. {\sl Proc. Natl. Acad. Sci. USA} {\bf 93}: 9533-9538.\\

\noindent Irb{\" a}ck, A. and Sandelin, E. 2000. On Hydrophobicity
Correlations in Protein Chains. {\sl Biophys. J.} {\bf 79:}
2252-2258. \\

\noindent Kamtekar, S., Schiffer, J. M., Xiong, H., Babik,
J. M., and Hecht, M. H. 1993. Protein design by binary patterning of polar and 
nonpolar amino-acids. {\sl Science} {\bf 262:} 1680-1685.\\

\noindent Kauzmann, W. 1959. Some factors in the interpretation of
protein denaturation. {\sl Adv. Protein Chem.} {\bf 14:} 1-63. \\

\noindent Kyte, J., and Doolittle, R.F. 1982. A simple method for
displaying the hydropathid character of a protein. {\sl J. Mol. Biol.}
{\bf 157:} 105-132. \\

\noindent Lau, K.F. and Dill, K.A. 1989 A lattice statistical
mechanics model of the conformational and sequence spaces if
proteins. {\sl Macromolecules} {\bf 22:} 3986-3997. \\

\noindent Lee, B. 1993. Estimation of the Maximum Change in Stability
of Globular Proteins upon Mutation of a Hydrophobic Residue to another
of Smaller Size. {\sl Prot. Sci.} {\bf 2:} 733-738.\\

\noindent Lesser, G.J. and Rose, G.D. 1990. Hydrophobicity of amino
acid subgroups in proteins. {\sl Proteins} {\bf 8:} 6-13. \\

\noindent Levitt, M. and Chothia, C. 1976. Structural patterns in
globular proteins. {\sl Nature} {\bf 261:} 552-558. \\

\noindent Li, H., Helling, R., Tang, C., and Wingreen,
N. 1996. Emergence of preferred structures in a simple model of
protein folding. {\sl Science} {\bf 273:} 666-669.  \\

\noindent Li, H., Tang, C., and Wingreen, N. 1998. Are protein folds
atypical? {\sl Proc. Natl. Acad. Sci.} USA {\bf 95:} 4987-4990. \\

\noindent Lins, L., Thomas, A., and Brasseur, R. 2003. Analysis of
accessible surface of residues in proteins. {\sl Protein Sci.} {\bf
12:} 1406-1417. \\

\noindent Miller, J., Zeng, C., Wingreen, N., and Tang,
C. 2002. Emergence of highly-designable protein-backbone conformations
in an off-lattice model. {\sl Proteins} {\bf 47:} 506-512. \\

\noindent Miller, S., Janin, J., Lesk, A.M., and Chothia,
C. 1987. Interior and surface of monomeric proteins. {\sl
J. Mol. Biol.} {\bf 196:} 641-56. \\

\noindent Miyazawa, S. and Jernigan, R.L. 1996. Residue-residue
potentials with a favorable contact pair term and an unfavorable high
packing term, for simulation and threading. {\sl J. Mol. Biol.} {\bf
256:} 623-644. \\

\noindent Miyazawa, S. and Jernigan, R.L. 1999. Self-consistent
estimation of inter-residue protein contact energies based on an
equilibrium mixture approximation of residues. {\sl Proteins:
Structures and Molecular Principles} {\bf 34:} 49-68. \\

\noindent Munoz, V. Serrano, L. 1994. Intrinsic secondary structure
propensities of the amino acids, using statistical phi-psi matrices:
comparison with experimental scales. {\sl Proteins} {\bf 20}:
301-311.\\

\noindent Murzin, A.G., Brenner, S.E., Hubbard, T., Chothia,
C. 1995. SCOP: a structural classification of proteins database for
the investigation of sequences and structures. {\sl J. Mol. Biol.}
{\bf 247:} 536-540. http://scop.mrc-lmb.cam.ac.uk/scop/ \\

\noindent Nauchitel, V.V. and Somorjai, R.L. 1994. Spatial and free
energy distribution patterns of amino acid residues in water soluble
proteins. {\sl Biophys. Chem.} {\bf 51:} 327-336. \\

\noindent Nozaki, Y. and Tanford, C. 1971. Solubility of Amino Acids
and 2 Glycine Peptides in Aqueous Ethanol AND Dioxane Solutions -
Establishment of a Hydrophobicity Scale. {\sl J. Bio. Chem} {\bf 246:}
2211. \\

\noindent Rose, G. Geselowitz, A., Lesser, G., Lee, R., and Zehfus,
M. 1985. Hydrophobicity of Amino Acid Residues in Globular
Proteins. {\sl Science} {\bf 289:} 834-839. \\

\noindent Tanford, C. 1978. Hydrophobic Effect and Organization of
Living Matter. {\sl Science} {\bf 200:} 1012-1018. \\

\noindent Taverna, D. and Goldstein, R.A.  2002a. Why are proteins
marginally stable? {\sl Proteins} {\bf 46:} 105-109. \\

\noindent Taverna, D.M. and Goldstein, R.A. 2002b. Why are proteins so
robust to site mutations? {\sl J. Mol. Bio.} {\bf 315}:479-484. \\

\noindent White, S.~H. and Jacobs, R.~E. 1990. Statistical
distribution of hydrophobic residues along the length of protein
chains. Implications for protein folding and evolution. {\sl
Biophys. J.} {\bf 57}: 911-921.\\

\begin{center}
\begin{table}[h]
\includegraphics[width=10cm]{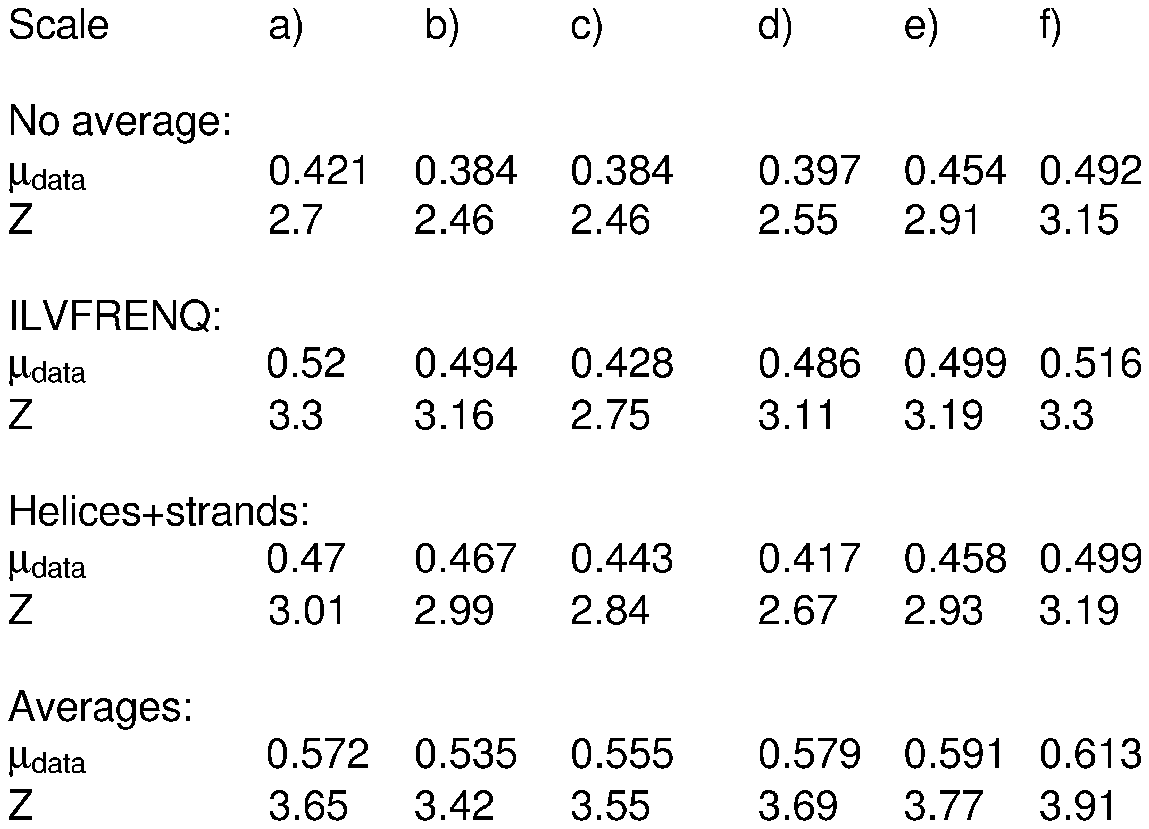}
\vspace{1in}
\caption{Summary of correlation analysis. Scales used are: a) Kyte and
Doolittle (1982), b) Engelman et al.  (1986), c) Nozaki and Tanford (1993), d)
Miyazawa and Jernigan (1996) and e) Miyazawa and Jernigan (1999), and
f) ASA. The mean correlation coefficient ($\mu_{data}$) of each
distribution is given along with the $Z =
(\mu_{data}-\mu_{random})/\sigma_{random}$ for several different
conditions. No average corresponds to using just individual sequences
and structures. ILVFRENQ considered only those positions with the
given amino acids. Helices+strands used only those residues that
formed secondary structural elements. Finally averages computed the
correlation coefficient using an average sequence computed from the
set of aligned sequences for a given representative structure.
\label{Tab1}}
\end{table}
\end{center}

\begin{center}
\begin{table}[h]
\includegraphics[width=15cm]{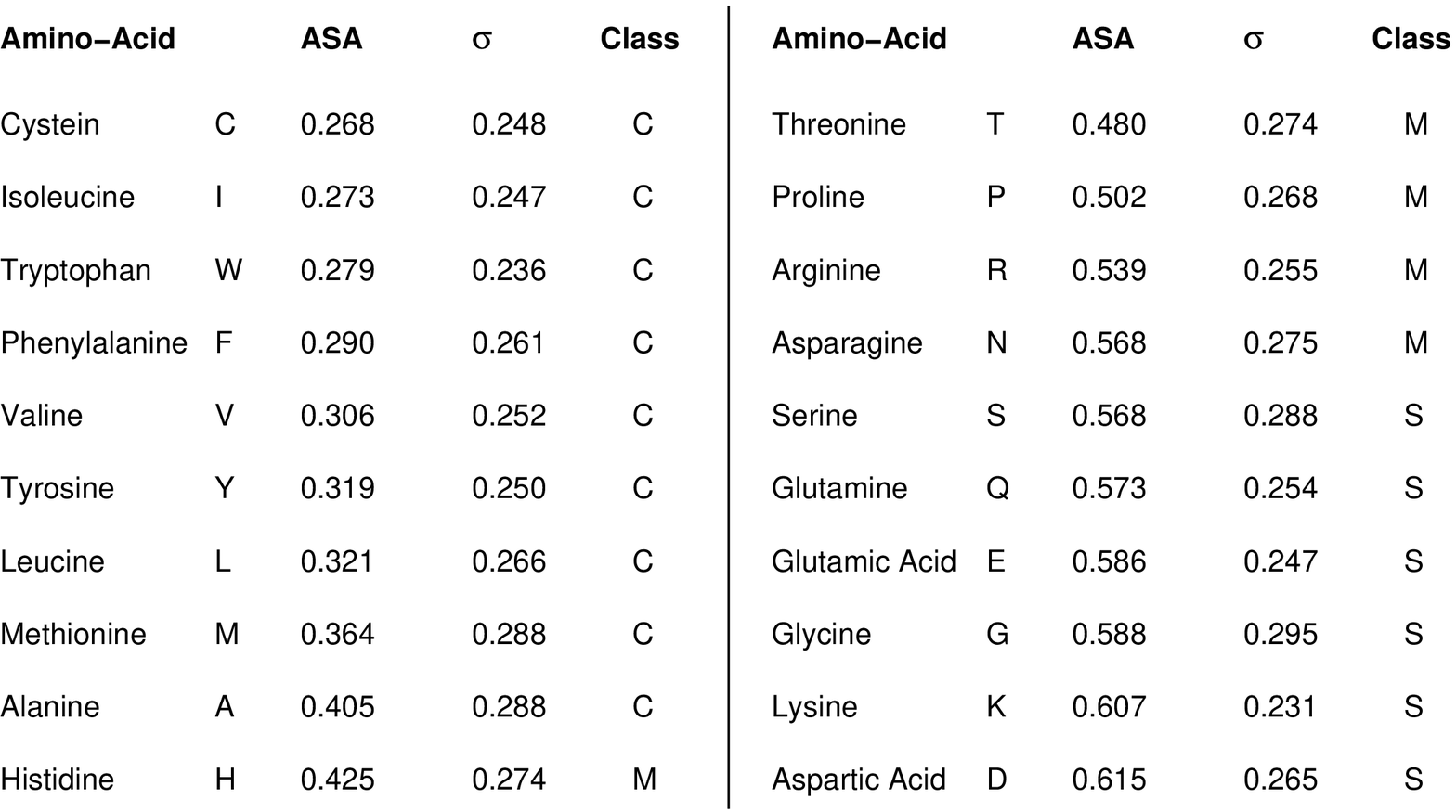}
\vspace{1in}
\caption{ASAs of amino-acids obtained by analysis of the complete
structure and sequence database, and their classification based on 
surface-accessibility distribution 
(Figs.~\ref{Fig4}-\ref{Fig6}). The variances, $\sigma$, of each distribution are also given  }
\label{Tab2}
\end{table}
\end{center}

\end{document}